\documentclass[12pt]{article}
\usepackage{amsmath}
\usepackage{graphicx}
\usepackage{amsfonts}
\usepackage{amssymb}
\input epsf

\begin{document}
\begin{flushright}
\setlength{\baselineskip}{3ex}
HUTP-99/A003
\end{flushright}

\begin{center}
{\Large \textbf{Renormalization in spherical field theory}}$\footnote{Support
provided by the NSF under Grant 5-22968 and PHY-9802709}$

\bigskip

\textrm{Dean Lee}\footnote{email: dlee@het.phast.umass.edu}

University of Massachusetts

Amherst, MA 01003

\bigskip\textrm{Nathan Salwen}\footnote{email: salwen@physics.harvard.edu}

Harvard University

Cambridge, MA 02138

{\Large {\normalsize \textrm{\newline \bigskip\ {\small \parbox
{360pt}{We derive several results concerning non-perturbative renormalization
in the spherical field formalism.  Using a small set of
local counterterms, we are able to remove all ultraviolet divergences
in a manner such that the renormalized theory is finite and
translationally invariant.  As an  explicit example  we consider massless
$\phi^{4}%
$  theory in four dimensions. [PACS numbers: 11.10.-z, 11.10.Gh, 11.15.Tk] }}%
}}}
\end{center}

\section{{\protect\Large {\protect\normalsize \textrm{{\protect\small \vspace
{10pt} }}}}Introduction}

Spherical field theory is a non-perturbative method which uses the spherical
partial wave expansion to reduce a general $d$-dimensional Euclidean field
theory into a set of coupled radial systems (\cite{a1}$,$\ \cite{a2}). High
spin partial waves correspond with large tangential momenta and can be
neglected if the theory is properly renormalized. The remaining system can
then be converted into differential equations and solved using standard
numerical methods. $\phi^{4}$ theory in two dimensions was considered in
\cite{a1}. In that case there was only one divergent diagram, and it could be
completely removed by normal ordering. In general any super-renormalizable
theory can be renormalized by removing the divergent parts of divergent
diagrams. Using a high-spin cutoff $J_{\max}$ and discarding partial waves
with spin greater than $J_{\max}$, we simply compute the relevant counterterms
using spherical Feynman rules.

The $J_{\max}$ cutoff scheme however is not translationally invariant. It
preserves rotational invariance but regulates ultraviolet processes
differently depending on radial distance. In the two-dimensional $\phi^{4}$
example it was found that the mass counterterm had the form
\begin{equation}
\mathcal{L}_{c.t.}\propto\phi^{2}(\vec{t})\left[  K_{0}(\mu t)I_{0}(\mu t)+2%
{\textstyle\sum_{n=1,J_{\max}}}
K_{n}(\mu t)I_{n}(\mu t)\right]  ,
\end{equation}
where $I_{n}$, $K_{n}$ are $n^{\text{th}}$-order modified Bessel functions of
the first and second kinds, $\mu$ is the bare mass, and $t$ is the magnitude
of $\vec{t}$. As $J_{\max}\rightarrow\infty$, we find%
\begin{equation}
\mathcal{L}_{c.t.}\propto\phi^{2}(\vec{t})\left[  \log(\tfrac{2J_{\max}}{\mu
t})+O(J_{\max}^{-1})\right]  .
\end{equation}
Our regularization scheme varies with $t$, and we see that the counterterm
also depends on $t$. The physically relevant issue, however, is whether or not
the renormalized theory is independent of $t$. In this case the answer is yes.
Any $t$ dependence in renormalized amplitudes is suppressed by powers of
$J_{\max}^{-1}$, and translational invariance becomes exact as $J_{\max
}\rightarrow\infty$.

We now consider general renormalizable theories, in particular those which are
not super-renormalizable. In this case the number of divergent diagrams is
infinite. Since we are primarily interested in non-perturbative phenomena, a
diagram by diagram subtraction method is not useful. In the same manner
strictly perturbative methods such as dimensional regularization are not
relevant either. Our interest is in non-perturbative renormalization, where
coefficients of renormalization counterterms are determined by
non-perturbative computations.\footnote{We should mention that Pauli-Villars
regularization is compatible with non-perturbative renormalization. \ However
this introduces additional unphysical degrees of freedom and tends to be
computationally inefficient.} In this paper we analyze the general theory of
non-perturbative renormalization in the spherical field formalism. In the
course of our analysis we answer the following three questions: (i) Can
ultraviolet divergences be cancelled by a finite number of local counterterms?
(ii) Can the renormalized theory be made translationally invariant? (iii) What
is the general form of the counterterms?

The organization of the paper is as follows. We begin with a discussion of
differential renormalization, a regularization-independent method which will
allow us to construct local counterterms. Next we describe a regularization
procedure which is convenient for spherical field theory. In the large radius
limit $t\rightarrow\infty$ our regularization procedure (which we call angle
smearing) is anisotropic but locally invariant under translations. For general
$t$ we expand in powers of $t^{-1}$ to generate the general form of the
counterterms. We conclude with two examples of one-loop divergent diagrams. We
show by direct calculation that the predicted counterterms render these
processes finite and translationally invariant.

\section{Differential renormalization}

Differential renormalization is the coordinate space version of the BPHZ
method.\footnote{Paraphrase of private communication with Jose Latorre.} It is
framed entirely in coordinate space, and renormalized amplitudes can be
defined as distributions without reference to any specific regularization
procedure. Differential renormalization was introduced in \cite{b}, and a
systematic analysis of differential renormalization to all orders in
perturbation theory using Bogoliubov's recursion formula was first described
in \cite{f}. The usual implementation of differential renormalization is
carried out using singular Poisson equations and their explicit solutions. In
our discussion, however, we find it more convenient to operate directly on the
distributions.\footnote{Our approach is similar to the natural renormalization
scheme described in \cite{d}. In contrast with \cite{d}, however, we do not a
priori specify the finite parts of amplitudes.} We describe the details of our
approach in the following. We should stress that the two approaches are
equivalent, differing only at the level of formalism.

We assume that we are working with a renormalizable theory. For indices
$i_{1},\cdots i_{j}$ let us define%
\begin{align}
t^{i_{1},\cdots i_{j}}  &  =t^{i_{1}}t^{i_{2}}\cdots t^{i_{j}},\\
\nabla_{i_{1},\cdots i_{j}}  &  =\nabla_{i_{1}}\nabla_{i_{2}}\cdots
\nabla_{i_{j}}.
\end{align}
Let $f(\vec{t})$ be a smooth test function, and let $I(\vec{t}-\vec{t}%
^{\prime};\mu^{2})$ be a smooth function with support on a region of scale
$\mu^{-1}$. We define $S_{\vec{t}^{\prime}}^{j}\left[  f\right]  (\vec{t})$ as
$I(\vec{t}-\vec{t}^{\prime};\mu^{2})$ multiplied by the $j^{th}$ order term in
the Taylor series of $f(\vec{t})$ about the point $\vec{t}^{\prime}$.
Inserting delta functions, we have
\begin{align}
S_{\vec{t}^{\prime}}^{j}f(\vec{t})  &  =I(\vec{t}-\vec{t}^{\prime};\mu^{2})%
{\textstyle\sum\limits_{i_{1},\cdots i_{j}}}
\left[  \tfrac{(t-t^{\prime})^{i_{1},\cdots i_{j}}}{j!}\nabla_{i_{1},\cdots
i_{j}}f(\vec{t}^{\prime})\right] \\
&  =I(\vec{t}-\vec{t}^{\prime};\mu^{2})%
{\textstyle\sum\limits_{i_{1},\cdots i_{j}}}
\tfrac{(t-t^{\prime})^{i_{1},\cdots i_{j}}}{j!}%
{\textstyle\int}
d^{4}\vec{z}\,\,\nabla_{i_{1},\cdots i_{j}}^{\vec{t}^{\prime}}\delta^{4}%
(\vec{t}^{\prime}-\vec{z})\,f(\vec{z}).\nonumber
\end{align}
For the purposes of this discussion we will require%
\begin{equation}
I(\vec{t}-\vec{t}^{\prime};\mu^{2})=1+O^{N}(\vec{t}-\vec{t}^{\prime
})\text{\quad as }\vec{t}^{\prime}\rightarrow\vec{t}\text{,}%
\end{equation}
where $N$ is some positive integer greater than the superficial degree of
divergence of any subdiagram\footnote{In our discussion a subdiagram is a
subset of vertices together with all lines contained in those vertices.} in
the theory we are considering. \ For any renormalizable theory $N>2$ will
suffice. In our formalism, $I(\vec{t}-\vec{t}^{\prime};\mu^{2})$ determines
how finite parts of renormalized amplitudes are assigned, and $\mu$ is the
renormalization mass scale.

We now consider a particular diagram, $G$, with $n$ vertices. We define
$K(\vec{t}_{1},\cdots\vec{t}_{n})$ to be the kernel of the amputated diagram,
i.e., the value of the diagram with vertices fixed at points $\vec{t}%
_{1},\cdots\vec{t}_{n}$. The amplitude is obtained by integrating $K(\vec
{t}_{1},\cdots\vec{t}_{n})$\ with respect to all internal vertices. We will
regard $K$ as a distribution acting on $n$ smooth test functions $f_{1},\cdots
f_{n}.\ $(For external vertices containing more than one external line and/or
derivatives, $f_{ext}(\vec{t}_{ext})$ should be regarded as a product of test
functions, with possible derivatives, at $\vec{t}_{ext}$.)%
\begin{equation}
K:f_{1},\cdots f_{n}\rightarrow%
{\textstyle\int}
d^{4}\vec{t}_{1}\cdots d^{4}\vec{t}_{n}\,K(\vec{t}_{1},\cdots\vec{t}_{n}%
)f_{1}(\vec{t}_{1})\cdots f_{n}(\vec{t}_{n}).
\end{equation}
Let us assume that our diagram is primitively divergent with superficial
degree of divergence $j$. We now define another distribution $T_{G}K$, which
extracts the divergent part of $K$. We start with the case when $G$ has more
than one vertex. Let us define $T_{G}K:f_{1},\cdots f_{n}\rightarrow$%

\begin{equation}%
{\textstyle\sum\limits_{_{j_{1}+\cdots+j_{n}\leq j}}}
{\textstyle\int}
d^{4}\vec{t}_{1}\cdots d^{4}\vec{t}_{n}\,K(\vec{t}_{1},\cdots\vec{t}%
_{n})S_{\vec{t}_{ave}}^{j_{1}}f_{1}(\vec{t}_{1})\cdots S_{\vec{t}_{ave}%
}^{j_{n}}f_{n}(\vec{t}_{n}),
\end{equation}
where $\vec{t}_{ave}=\tfrac{1}{n}(\vec{t}_{1}+\cdots+\vec{t}_{n}).$ We note
that the subtracted distribution $K-T_{G}K$ is finite and well-defined for all
$f_{1},\cdots f_{n}$. Let us define
\begin{align}
&  F_{K}^{i_{1,1},i_{2,1}\cdots i_{j_{n},n}}(\vec{t})\label{bw}\\
&  =%
{\textstyle\int}
d^{4}\vec{t}_{1}\cdots d^{4}\vec{t}_{n}\,\delta^{4}(\tfrac{\vec{t}_{1}%
+\cdots+\vec{t}_{n}}{n}-\vec{t})K(\vec{t}_{1},\cdots\vec{t}_{n})\left[
{\textstyle\prod\limits_{k=1,\cdots n}}
\tfrac{I(\vec{t}_{k}-\vec{t};\mu^{2})(t_{k}-t)^{i_{1,k},\cdots i_{j_{k},k}}%
}{j_{k}!}\right]  .\nonumber
\end{align}
We can then rewrite $T_{G}K:f_{1},\cdots f_{n}\rightarrow$%
\begin{equation}%
{\textstyle\sum\limits_{_{j_{1}+\cdots+j_{n}\leq j}}}
{\textstyle\sum\limits_{_{\substack{i_{1,1},i_{2,1}\cdots\\ i_{1,n}\cdots
i_{j_{n},n}}}}}
\left[
\begin{array}
[c]{c}%
{\textstyle\int}
d^{4}\vec{t}F_{K}^{i_{1,1},i_{2,1}\cdots i_{j_{n},n}}(\vec{t})\int d^{4}%
\vec{z}_{1}\cdots d^{4}\vec{z}_{n}\\
\left(  \prod_{k=1,\cdots n}\,\,\nabla_{i_{1,k},\cdots i_{j_{k},k}}^{\vec{t}%
}\delta^{4}(\vec{t}-\vec{z}_{k})\right)  f_{1}(\vec{z}_{1})\cdots f_{n}%
(\vec{z}_{n})
\end{array}
\right]  . \label{cou}%
\end{equation}
The delta functions make this kernel completely local. We can read off the
corresponding counterterm interaction by functional differentiation with
respect to each of the component functions of $f_{ext}(\vec{t}_{ext})$ for the
external vertices and setting $f_{int}(\vec{t}_{int})=1$ for the internal
vertices. We now turn to the case when $G$ has only one vertex. For this case
we set $T_{G}K=K$, which is equivalent to normal ordering the interactions in
our theory.\ In this case $K$ is itself local and therefore $T_{G}K$ and our
counterterm interaction are again local.

We now extend the definition of $T_{G}$ in (\ref{cou}) to include the case of
subdiagrams. Let $G$ be a general 1PI diagram, and let $G^{\prime}$ be a
renormalization part\footnote{A renormalization part is a 1PI subdiagram with
degree of divergence $\geq0$.} of $G$ with superficial degree of divergence
$j^{\prime}$. For notational convenience we will label the vertices of $G$ so
that the first $n^{\prime}$ vertices lie in $G^{\prime}$. If $G^{\prime}$ has
only one vertex then again we normal order the interaction. Otherwise we
define $T_{G^{\prime}}K:f_{1},\cdots f_{n}\rightarrow$
\begin{equation}%
{\textstyle\sum\limits_{j_{1}^{\prime}+\cdots+j_{n^{\prime}}^{\prime}\leq
j^{\prime}}}
{\textstyle\int}
d^{4}\vec{t}_{1}\cdots d^{4}\vec{t}_{n}\,K(\vec{t}_{1},\cdots\vec{t}%
_{n})\left[
\begin{array}
[c]{c}%
S_{\vec{t}_{ave}}^{j_{1}^{\prime}}f_{1}(\vec{t}_{1})\cdots S_{\vec{t}_{ave}%
}^{j_{n}^{\prime}}f_{n^{\prime}}(\vec{t}_{n^{\prime}})\\
\cdot f_{n^{\prime}+1}(\vec{t}_{n^{\prime}+1})\cdots f_{n}(\vec{t}_{n})
\end{array}
\right]  ,
\end{equation}
where $\vec{t}_{ave}=\tfrac{1}{n^{\prime}}(\vec{t}_{1}+\cdots+\vec
{t}_{n^{\prime}})$.\footnote{After applying $T_{G^{\prime}}$, we regard
$G^{\prime}$ as being contracted to single vertex at $\vec{t}_{ave}$.} This
definition can be used recursively to define products of $T_{G_{1}^{\prime}%
}T_{G_{2}^{\prime}}$ for disjoint subdiagrams $G_{1}^{\prime}\cap
G_{2}^{\prime}=\emptyset$ or nested subdiagrams $G_{1}^{\prime}\supset
G_{2}^{\prime}.$ For the case of nested subdiagrams we always order the
product so that larger diagrams are on the left.

It is straightforward to show that the $T$ operation acts as the identity on
local interactions and thus treats overlapping divergences in the same manner
as BPHZ. Following the standard BPHZ procedure (\cite{g}$-$\cite{i}), we can
write Bogoliubov's $\bar{R}$ operation using Zimmerman's forest formula,%
\begin{equation}
\bar{R}=%
{\textstyle\sum\limits_{F}}
{\textstyle\prod\limits_{\gamma\in F}}
(-T_{\gamma}),
\end{equation}
where $F$ ranges over all forests\footnote{A forest is any set of
non-overlapping renormalization parts.} of $G$, and $\gamma$ ranges over all
renormalization parts of $F.$ In the product we have again ordered nested
subdiagrams so that larger diagrams are on the left. Let $j$ be the
superficial degree of divergence of $G$. The renormalized kernel$,$ $RK$, is
given by%
\begin{equation}
\left.
\begin{array}
[c]{c}%
RK=\bar{R}K\quad\\
RK=(1-T_{G})\bar{R}K\quad
\end{array}
\right.  \left.
\begin{array}
[c]{c}%
\text{for }j<0\\
\text{for }j\geq0.
\end{array}
\right.
\end{equation}
Our final result is that all required counterterms are local, and the form of
the counterterms is
\begin{equation}
\mathcal{L}_{c.t.}=%
{\textstyle\sum_{A(\phi,\nabla_{i}\phi)\text{ }}}
F_{A}(\vec{t})A(\phi(\vec{t}),\nabla_{i}\phi(\vec{t})), \label{be}%
\end{equation}
where the sum is over operators of renormalizable type. For the case of gauge
theories, our renormalization procedure is supplemented by the additional
requirement that the renormalized amplitudes satisfy Ward
identities.\footnote{See \cite{e1}, \cite{e2} for a discussion of gauge
theories using the method of differential renormalization.} If our
regularization procedure breaks gauge invariance these identities are not
automatic and the required local counterterms will in general be any operators
of renormalizable type (not merely gauge-invariant operators). This is,
however, a separate discussion, and the details of implementing Ward identity
constraints will be discussed in future work.

\section{\bigskip Regularization by angle smearing}

In this section we determine the functional form of the coefficients
$F_{A}(\vec{t})$ in (\ref{be}). To make the discussion concrete, we will
illustrate using the example of massless $\phi^{4}$ theory in four dimensions
\begin{equation}
\mathcal{L}=\tfrac{1}{2}\phi\nabla^{2}\phi-\tfrac{\lambda}{4!}\phi
^{4}+\mathcal{L}_{c.t.}.
\end{equation}
From (\ref{be}) $\mathcal{L}_{c.t.}$ is given by%
\begin{equation}
F_{\phi^{2}}(\vec{t})\phi^{2}(\vec{t})+\text{ }%
{\textstyle\sum_{i,j}}
F_{\nabla\phi\nabla\phi}^{ij}(\vec{t})\nabla_{i}\phi(\vec{t})\nabla_{j}%
\phi(\vec{t})+F_{\phi^{4}}(\vec{t})\phi^{4}(\vec{t}).\text{ }%
\end{equation}
Let $G(\vec{t},\vec{t}^{\prime})$ be the free two-point correlator. We will
use a regularization scheme which preserves rotational invariance and is
convenient for spherical field theory, but one which breaks translational
invariance. We regulate the short distance behavior of $G$ by smearing the
endpoints over a radius $t$ spherical shell within a conical region $R_{M^{2}%
}(\vec{t})$, where $R_{M^{2}}(\vec{t})$ is the set of vectors $\vec{u}$ such
that the angle between $\vec{t}$ and $\vec{u}$ is between $-\frac{1}{Mt}$ and
$\frac{1}{Mt}$ (see Figure 1). The result is a regulated correlator
\begin{equation}
G_{M^{2}}(\vec{t},\vec{t}^{\prime})=\tfrac{1}{\int_{\hat{u}\in R_{M^{2}}%
(\vec{t})}d^{3}\hat{u}\int_{\hat{u}^{\prime}\in R_{M^{2}}(\vec{t}^{\prime}%
)}d^{3}\hat{u}^{\prime}\,\,}\int_{\substack{\hat{u}\in R_{M^{2}}(\vec
{t})\\\hat{u}^{\prime}\in R_{M^{2}}(\vec{t}^{\prime})}}d^{3}\hat{u}d^{3}%
\hat{u}^{\prime}\,G(t\hat{u},t^{\prime}\hat{u}^{\prime}). \label{as}%
\end{equation}
We recall that our renormalized theory is determined by the translationally
invariant function $I(\vec{t}-\vec{t}_{ave};\mu^{2})$ described in the
previous section. Even though our regularization scheme breaks translational
invariance, the renormalized theory nevertheless remains invariant.

As the radius $t$ increases the curvature of the angle-smearing region becomes
negligible. In the limit $t\rightarrow\infty$ the region becomes a flat
three-dimensional ball with radius $\tfrac{1}{M}$ lying in the plane
perpendicular to the radial vector. In this limit our regularization is
invariant under local transformations and the counterterms converge to
constants independent of $\vec{t}$,%
\begin{align}
\lim_{t\rightarrow\infty}F_{\nabla\phi\nabla\phi}^{ij}(\vec{t})  &
=c_{\nabla\phi\nabla\phi}^{ij,(0)}(\tfrac{\mu^{2}}{M^{2}})\\
\lim_{t\rightarrow\infty}F_{\phi^{2}}(\vec{t})  &  =M^{2}c_{\phi^{2}}%
^{(0)}(\tfrac{\mu^{2}}{M^{2}})\\
\lim_{t\rightarrow\infty}F_{\phi^{4}}(\vec{t})  &  =c_{\phi^{4}}^{(0)}%
(\tfrac{\mu^{2}}{M^{2}}).
\end{align}
We have chosen our coefficients $c_{A}^{(0)}$ to be dimensionless. Although
our regularization scheme is invariant under rotations about the origin, the
radial vector has a special orientation which is normal to our
three-dimensional ball. Our regularization scheme is therefore not isotropic.
The result (as should be familiar from studies of anisotropic lattices) is
that the coefficient of the kinetic term has two independent components%
\begin{equation}
c_{\nabla\phi\nabla\phi}^{ij,(0)}(\tfrac{\mu^{2}}{M^{2}})=c_{\nabla\phi
\nabla\phi}^{\hat{t}\hat{t},(0)}(\tfrac{\mu^{2}}{M^{2}})+\delta^{ij}%
c_{\nabla\phi\nabla\phi}^{(0)}(\tfrac{\mu^{2}}{M^{2}}).
\end{equation}

Starting with the $t\rightarrow\infty$ result at lowest order, we now expand
our coefficient functions in powers of $\frac{1}{Mt}$,%
\begin{align}
F_{\nabla\phi\nabla\phi}^{ij}(\vec{t})  &  =c_{\nabla\phi\nabla\phi}%
^{ij,(0)}(\tfrac{\mu^{2}}{M^{2}})+\tfrac{1}{Mt}c_{\nabla\phi\nabla\phi
}^{ij,(1)}(\tfrac{\mu^{2}}{M^{2}})+\tfrac{1}{M^{2}t^{2}}c_{\nabla\phi
\nabla\phi}^{ij,(2)}(\tfrac{\mu^{2}}{M^{2}})+\cdots\\
F_{\phi^{2}}(\vec{t})  &  =M^{2}c_{\phi^{2}}^{(0)}(\tfrac{\mu^{2}}{M^{2}%
})+\tfrac{M}{t}c_{\phi^{2}}^{(1)}(\tfrac{\mu^{2}}{M^{2}})+\tfrac{1}{t^{2}%
}c_{\phi^{2}}^{(2)}(\tfrac{\mu^{2}}{M^{2}})+\cdots\\
F_{\phi^{4}}(\vec{t})  &  =c_{\phi^{4}}^{(0)}(\tfrac{\mu^{2}}{M^{2}}%
)+\tfrac{1}{Mt}c_{\phi^{4}}^{(1)}(\tfrac{\mu^{2}}{M^{2}})+\tfrac{1}{M^{2}%
t^{2}}c_{\phi^{4}}^{(2)}(\tfrac{\mu^{2}}{M^{2}})+\cdots.
\end{align}
For the moment let us assume\ $t\geq\Lambda^{-1}$ for%
\begin{equation}
\Lambda=m_{0}^{z}M^{1-z},
\end{equation}
for some fixed mass $m_{0}$ and constant $z$ such that $0<z<\frac{1}{2}$. In
this region our dimensionless expansion parameter $\frac{1}{Mt}$ is bounded by
$\tfrac{m_{0}^{z}}{M^{z}}$ and therefore diminishes uniformly as
$M\rightarrow\infty$.

In general the $\tfrac{\mu^{2}}{M^{2}}$ dependence in the functions
$c_{A}^{(k)}$ will contain analytic terms as $\mu^{2}\rightarrow0$ as well as
logarithmically divergent terms. There are, however, no inverse powers of
$\tfrac{\mu^{2}}{M^{2}}$. These would indicate severe infrared divergences not
present in the processes we are considering, as can be deduced from the long
distance behavior of the integral in (\ref{bw}).\footnote{If our theory
contained bare masses $m_{i}$, similar arguments would apply for the infrared
limit $\mu^{2},m_{i}^{2}\rightarrow0,$ with $\tfrac{m_{i}^{2}}{\mu^{2}}$
fixed.} With this we can neglect terms which vanish as $M\rightarrow\infty,$
\begin{align}
F_{\phi^{2}}(\vec{t})  &  =M^{2}c_{\phi^{2}}^{(0)}(\tfrac{\mu^{2}}{M^{2}%
})+\tfrac{1}{t^{2}}c_{\phi^{2}}^{(2)}(\tfrac{\mu^{2}}{M^{2}})\\
F_{\nabla\phi\nabla\phi}^{ij}(\vec{t})  &  =c_{\nabla\phi\nabla\phi}%
^{ij,(0)}(\tfrac{\mu^{2}}{M^{2}})\\
F_{\phi^{4}}(\vec{t})  &  =c_{\phi^{4}}^{(0)}(\tfrac{\mu^{2}}{M^{2}}).
\end{align}
Since our regularization scheme is invariant under $M\rightarrow-M$, we have
also omitted the term proportional$\ $to $c_{\phi^{2}}^{(1)}$ which is odd in
$M$.

We now consider what occurs in the small region near the origin, $t\leq
\Lambda^{-1}$. For the theory we are considering (and in fact for any
renormalizable theory) the highest ultraviolet divergence possible is
quadratic.\footnote{There may be additional logarithmic factors but this does
not matter for our purposes here.} In the limit $M\rightarrow\infty$ we deduce
that each $F_{A}$ scales no greater than $O(M^{2}).$ On the other hand the
volume of the region $t\leq\Lambda^{-1}$ diminishes as $O(M^{4z-4}).$ Thus the
total contribution from the region $t\leq\Lambda^{-1}$ scales as $O(M^{4z-2})$
and can be entirely neglected.

To summarize our results, the counterterm Lagrange density has the form
\begin{equation}
c_{\nabla\phi\nabla\phi}^{(0)}(\vec{\nabla}\phi(\vec{t}))^{2}+c_{\nabla
\phi\nabla\phi}^{\hat{t}\hat{t},(0)}(\hat{t}\cdot\vec{\nabla}\phi(\vec
{t}))^{2}+(M^{2}c_{\phi^{2}}^{(0)}+\tfrac{1}{t^{2}}c_{\phi^{2}}^{(2)})\phi
^{2}(\vec{t})+c_{\phi^{4}}^{(0)}\phi^{4}(\vec{t}). \label{an}%
\end{equation}

\section{\bigskip Spherical fields}

We now examine the results of the previous section in the context of spherical
field theory. We start with the spherical partial wave expansion,%
\begin{equation}
\phi=%
{\textstyle\sum_{l=0,1,\cdots}}
{\textstyle\sum_{n=0,\cdots l}}
{\textstyle\sum_{m=-n,\cdots n}}
\phi_{l,n,m}(t)Y_{l,n,m}(\theta,\psi,\varphi),
\end{equation}
where $Y_{l,m,n}$ are four-dimensional spherical harmonics satisfying%
\begin{equation}%
{\textstyle\int}
d^{3}\Omega\,Y_{l^{\prime},n^{\prime},m^{\prime}}^{\ast}(\theta,\psi
,\varphi)Y_{l,n,m}(\theta,\psi,\varphi)=\delta_{l^{\prime},l}\delta
_{n^{\prime},n}\delta_{m^{\prime},m},
\end{equation}%
\begin{equation}
Y_{l,n,m}^{\ast}(\theta,\psi,\varphi)=(-1)^{m}Y_{l,n,-m}(\theta,\psi,\varphi).
\end{equation}
The explicit form of $Y_{l,m,n}$ can be found in \cite{j}.\footnote{\cite{j}
deserves credit as the first discussion of radial (or covariant Euclidean)
quantization, an important part of the spherical field formalism.} The
integral of the free massless Lagrange density in terms of spherical fields is%
\begin{equation}%
{\textstyle\int}
d^{4}\vec{t}\,\mathcal{L}=%
{\textstyle\int_{0}^{\infty}}
dt\,\left\{  \sum_{l,m,n}\left[  (-1)^{m}\phi_{l,n,-m}\left[  \tfrac{\partial
}{\partial t}\tfrac{t^{3}}{2}\tfrac{\partial}{\partial t}-\tfrac{t}%
{2}l(l+2)\right]  \phi_{l,n,m}\right]  \right\}  .
\end{equation}
It can be shown that the process of angle smearing the field $\phi(\vec{t})$
is equivalent to multiplying $\phi_{l,n,m}(t)$ by an extra factor $s_{l}%
^{M}(t)$ where%
\begin{equation}
s_{l}^{M}(t)=\tfrac{2Mt\left[  (l+2)\sin(\frac{l}{Mt})-l\sin(\frac{l+2}%
{Mt})\right]  }{l(l+1)(l+2)\left[  2-Mt\sin(\frac{2}{Mt})\right]  }.
\end{equation}
For large $l$, $s_{l}^{M}(t)$ diminishes as $l^{-2}$, and so the correlator
receives an extra suppression of $l^{-4}$. We will later use this result to
estimate the contribution of high spin partial waves. The regularization of
our correlator can be implemented in our Lagrange density by dividing factors
of $s_{l}^{M}(t),$
\begin{align}
&  \phi_{l,n,-m}\left[  \tfrac{\partial}{\partial t}\tfrac{t^{3}}{2}%
\tfrac{\partial}{\partial t}-\tfrac{t}{2}l(l+2)\right]  \phi_{l,n,m}\\
&  \rightarrow\left[  (s_{l}^{M}(t))^{-1}\phi_{l,n,-m}\right]  \left[
\tfrac{\partial}{\partial t}\tfrac{t^{3}}{2}\tfrac{\partial}{\partial
t}-\tfrac{t}{2}l(l+2)\right]  \left[  (s_{l}^{M}(t))^{-1}\phi_{l,n,m}\right]
.\nonumber
\end{align}
We now include the interaction and counterterms. We first define%
\begin{align}
&  \left[
\genfrac{}{}{0pt}{1}{l_{1},n_{1},m_{1};l_{2},n_{2},m_{2}}{l_{3},n_{3}%
,m_{3};l_{4},n_{4},m_{4}}%
\right] \\
&  =%
{\textstyle\int}
d^{3}\Omega\,Y_{l_{1},n_{1},m_{1}}(\theta,\psi,\varphi)Y_{l_{2},n_{2},m_{2}%
}(\theta,\psi,\varphi)Y_{l_{3},n_{3},m_{3}}(\theta,\psi,\varphi)Y_{l_{4}%
,n_{4},m_{4}}(\theta,\psi,\varphi).\nonumber
\end{align}
We can write the full functional integral as%
\begin{equation}
\int\mathcal{D}\phi\exp\left[
{\textstyle\int}
d^{4}\vec{t}\,\mathcal{L}\right]  \propto\int\left(
{\textstyle\prod_{l,n,m}}
\mathcal{D}\phi_{l,n,m}^{\prime}\right)  \exp\left[
{\textstyle\int_{0}^{\infty}}
dt\,(L_{1}+L_{2}+L_{3})\right]  ,
\end{equation}
where%
\begin{equation}
L_{1}=\sum_{l,m,n}\left[  (-1)^{m}\left[  (s_{l}^{M}(t))^{-1}\phi
_{l,n,-m}^{\prime}\right]  \left[  \tfrac{\partial}{\partial t}\tfrac{t^{3}%
}{2}\tfrac{\partial}{\partial t}-\tfrac{t}{2}l(l+2)\right]  \left[  (s_{l}%
^{M}(t))^{-1}\phi_{l,n,m}^{\prime}\right]  \right]  , \label{l1}%
\end{equation}%
\begin{equation}
L_{2}=\sum_{l,m,n}\left[  (-1)^{m}\phi_{l,n,-m}^{\prime}\left[
\begin{array}
[c]{c}%
\left[  -c_{\nabla\phi\nabla\phi}^{(0)}-c_{\nabla\phi\nabla\phi}^{\hat{t}%
\hat{t},(0)}\right]  \tfrac{\partial}{\partial t}\tfrac{t^{3}}{2}%
\tfrac{\partial}{\partial t}\\
+c_{\nabla\phi\nabla\phi}^{(0)}\tfrac{t}{2}l(l+2)+t^{3}(M^{2}c_{\phi^{2}%
}^{(0)}+\tfrac{1}{t^{2}}c_{\phi^{2}}^{(2)})
\end{array}
\right]  \phi_{l,n,m}^{\prime}\right]  , \label{l2}%
\end{equation}%
\begin{equation}
L_{3}=-t^{3}(\tfrac{\lambda}{4!}-c_{\phi^{4}}^{(0)})\sum_{l_{i},m_{i},n_{i}%
}\left[
\genfrac{}{}{0pt}{1}{l_{1},m_{1},n_{1};l_{2},m_{2},n_{2}}{l_{3},m_{3}%
,n_{3};l_{4},m_{4},n_{4}}%
\right]  \phi_{l_{1},m_{1},n_{1}}^{\prime}\phi_{l_{2},m_{2},n_{2}}^{\prime
}\phi_{l_{3},m_{3},n_{3}}^{\prime}\phi_{l_{4},m_{4},n_{4}}^{\prime}.
\label{l3}%
\end{equation}
We have used primes in preparation for redefining the fields,%
\begin{equation}
(s_{l}^{M}(t))^{-1}\phi_{l,n,m}^{\prime}=\phi_{l,n,m}.
\end{equation}
The Jacobian of this transformation is a constant (although infinite) and can
be absorbed into the normalization of the functional integral. Now the
Lagrangian $L_{1}$ has the usual free-field form in terms of $\phi_{l,n,m}$
while $L_{2}$ and $L_{3}$ are now functions of $s_{l}^{M}(t)\phi_{l,n,m}$.

With $M$ serving as our ultraviolet regulator, the contribution of
high-spin\ partial waves decouples for sufficiently large spin $l$. We can
estimate the order of magnitude of this contribution in the following manner.
We first identify $t^{-1}l$ (where $t$ is the characteristic radius we are
considering) as an estimate of the magnitude of the tangential momentum,
$p_{T}$. For $p_{T}\gg M\gg t^{-1}$ our correlator scales as $\tfrac{M^{4}%
}{p_{T}^{6}}.$ By dimensional analysis, a diagram with $N_{L}$ loops and
$N_{I}$ internal lines will receive a contribution from partial waves with
spin $\geq l$ of order%
\begin{equation}
\left(  \tfrac{M^{4}}{p_{T}^{6}}\right)  ^{N_{I}}\left(  p_{T}\right)
^{4N_{L}}=\left(  \tfrac{M^{4}}{(t^{-1}l)^{6}}\right)  ^{N_{I}}\left(
t^{-1}l\right)  ^{4N_{L}}. \label{est}%
\end{equation}
\qquad

\section{One-loop examples}

We will devote the remainder of our discussion to computing one-loop spherical
Feynman diagrams as a check of our results. Our calculations are done both
numerically and analytically. The diagrams we will consider are shown in
Figures 2 and 3. We start with the two-point function in Figure 2. The
amplitude can be written as $t^{3}B(t)$ where%
\begin{equation}
B(t)\propto%
{\textstyle\sum_{l,n,m}}
\tfrac{1}{t^{2}(l+1)}(s_{l}^{M}(t))^{2}. \label{g}%
\end{equation}
Constants of proportionality are not important here and so we will define
$B(t)$ to be equal to the right side of (\ref{g}). Our results tell us that if
we choose our mass counterterms appropriately, the combination
\begin{equation}
B(t)+M^{2}c_{\phi^{2}}^{(0)}+\tfrac{1}{t^{2}}c_{\phi^{2}}^{(2)}%
\end{equation}
should be independent of $t$, or more succinctly,%
\begin{equation}
B(t)+\tfrac{1}{t^{2}}c_{\phi^{2}}^{(2)}%
\end{equation}
is independent of $t$. Let us first check this analytically. In the absence of
a high-spin cutoff, we can explicitly calculate the sum in (\ref{g}):%
\begin{equation}
B(t)=\tfrac{1}{t^{2}}+b(t)
\end{equation}
where%
\begin{equation}
b(t)=\tfrac{4M^{2}\sin^{4}(\frac{1}{Mt})}{(2-Mt\sin(\frac{2}{Mt}))^{2}}.
\end{equation}
In the limit $M\rightarrow\infty,$%
\begin{equation}
B(t)\rightarrow\tfrac{1}{t^{2}}+\tfrac{9}{4}M^{2}.
\end{equation}
We conclude that $c_{\phi^{2}}^{(2)}=-1$ and $B(t)+\tfrac{1}{t^{2}}c_{\phi
^{2}}^{(2)}$ is in fact translationally invariant.

In Figure 4 we have plotted $B(t)-\tfrac{1}{t^{2}}$, computed numerically for
various values of the high-spin cutoff $J_{\max}$. We have also plotted the
limiting values $b(t)$ and $\tfrac{9}{4}M^{2}$. In our plot $t$ is measured in
units of $m^{-1}$ and $B(t)-\tfrac{1}{t^{2}}$ is in units of $m^{2}$, where
$m$ is an arbitrary mass scale such that $M=3m$. As expected, the errors are
of size $\frac{M^{4}t^{2}}{J_{\max}^{2}}$. There is clearly a deviation from
$\tfrac{9}{4}M^{2}$ for $t\lesssim M^{-1}$ but the integral of the deviation
is negligible as $M\rightarrow\infty$.

We now turn to the four-point function in Figure 3. The amplitude can be
written as $t_{1}^{3}t_{2}^{3}C(t_{1},t_{2})$ where%
\begin{equation}
C(t_{1},t_{2})\propto%
{\textstyle\sum_{l,n,m}}
\tfrac{(s_{l}^{M}(t_{1}))^{2}(s_{l}^{M}(t_{2}))^{2}}{(l+1)^{2}}\left[
\tfrac{t_{1}^{l}}{t_{2}^{l+2}}\theta(t_{2}-t_{1})+\tfrac{t_{2}^{l}}%
{t_{1}^{l+2}}\theta(t_{1}-t_{2})\right]  ^{2}. \label{sv}%
\end{equation}
Again constants of proportionally are not important and so we will define
$C(t_{1},t_{2})$ to be equal to the right side of (\ref{sv}). We can write
$C(t_{1},t_{2})$ in terms of the regulated correlator $G_{M^{2}}(\vec{t}%
_{1},\vec{t}_{2}),\footnote{We recall that the regulated correlator goes with
$\phi_{l,n,m}^{\prime}$ rather than $\phi_{l,n,m}$. But this is not important
here since $\phi_{0,0,0}^{\prime}$ $=\phi_{0,0,0}$.}$%
\begin{equation}
C(t_{1},t_{2})\propto\int d^{3}\hat{t}_{1}d^{3}\hat{t}_{2}\left[  G_{M^{2}%
}(\vec{t}_{1},\vec{t}_{2})\right]  ^{2}\propto\int d^{3}\hat{t}_{1}\left[
G_{M^{2}}(\vec{t}_{1},\vec{t}_{2})\right]  ^{2}.
\end{equation}
Since the coupling constant counterterm
\begin{equation}
c_{\phi^{4}}^{(0)}\delta^{4}(\vec{t}_{1}-\vec{t}_{2})
\end{equation}
is translationally invariant, the amplitude by itself should be
translationally invariant. Let us define
\begin{equation}
\int d^{4}\vec{t}_{2}e^{-i\vec{p}\cdot(\vec{t}_{1}-\vec{t}_{2})}\left[
G_{M^{2}}(\vec{t}_{1},\vec{t}_{2})\right]  ^{2}=f(\vec{p}^{2}),
\end{equation}
so that
\begin{equation}
\int d^{4}\vec{t}_{2}e^{i\vec{p}\cdot\vec{t}_{2}}\left[  G_{M^{2}}(\vec{t}%
_{1},\vec{t}_{2})\right]  ^{2}=e^{i\vec{p}\cdot\vec{t}_{1}}f(\vec{p}^{2}).
\end{equation}
Integrating over $\hat{t}_1$, we find%
\begin{equation}
\int dt_{2}\,t_{2}^{2}J_{1}(pt_{2})C(t_{1},t_{2})\propto\tfrac{1}{t_{1}}%
J_{1}(pt_{1})f(\vec{p}^{2}).
\end{equation}
Let us define%
\begin{equation}
C(t)=%
{\textstyle\int}
dt_{2}\,t_{2}^{2}J_{1}(pt_{2})C(t,t_{2}).
\end{equation}
We now check that in fact
\begin{equation}
C(t)\propto\tfrac{1}{t_{1}}J_{1}(pt_{1})\text{.}%
\end{equation}

In the absence of a high-spin cutoff, we find that $C(t)$ is given
by\footnote{This calculation is somewhat lengthy. Details can be obtained upon
request from the authors.}%
\begin{equation}
C(t)=\tfrac{1}{t_{1}}J_{1}(pt_{1})\left[  \tfrac{1}{2}\log\tfrac{M^{2}}{p^{2}%
}+c\right]  +\cdots,
\end{equation}
where the ellipsis represents terms which vanish as $M\rightarrow\infty$ and
\begin{equation}
c=324\left[  \int_{0}^{1/2}dk\left(  \tfrac{(\sin k-k\cos k)^{4}}{4k^{13}%
}-\tfrac{1}{324k}\right)  +\int_{1/2}^{\infty}dk\tfrac{(\sin k-k\cos k)^{4}%
}{4k^{13}}\right]  .
\end{equation}
In Figure 5 we plot $C(t)$ for different values of the high-spin cutoff
$J_{\max}$ as well as the large-$M$ limit value
\begin{equation}
C_{1}(t)=\tfrac{1}{t_{1}}J_{1}(pt_{1})\left[  \tfrac{1}{2}\log\tfrac{M^{2}%
}{p^{2}}+c\right]  .
\end{equation}
In our plot $t$ is measured in units of $p^{-1}$ and $M=3p$. From (\ref{est})
the expected error is of size $\frac{M^{8}t^{8}}{J_{\max}^{8}}.$ We see that
the data is consistent with the results expected. Again the deviation for
$t\lesssim M^{-1}$ integrates to a negligible contribution as $M\rightarrow
\infty$.

\section{Summary}

We have examined several important features of non-perturbative
renormalization in the spherical field formalism and answered the three
questions posed in the introduction. Ultraviolet divergences can be cancelled
by a finite number of local counterterms in a manner such that the
renormalized theory is translationally invariant. Using angle-smearing
regularization we find that the counterterms for $\phi^{4}$ theory in four
dimensions can be parameterized by five unknown constants as shown in
(\ref{an}). Aside from our remarks about Ward identity constraints in gauge
theories, the extension to other field theories is straightforward. We hope
that these results will be useful for future studies of general renormalizable
theories by spherical field techniques.\bigskip

\noindent{\Large \textbf{Acknowledgments}}\bigskip

\noindent We gratefully acknowledge useful correspondence with Daniel
Freedman, Roman Jackiw, and Jose Latorre.

\noindent{\Large \textbf{Figures}}\bigskip

\noindent Figure 1. Sketch of the angle-smearing region (three-dimensional rendering).

\noindent Figure 2. One-loop two-point correlator for $\phi_{0,0,0}$.

\noindent Figure 3. One-loop four-point correlator for $\phi_{0,0,0}$.

\noindent Figure 4. Plot of $B(t)-\tfrac{1}{t^{2}}$.

\noindent Figure 5. Plot of $C(t)$.

\begin{figure}[ptb]
\epsfbox{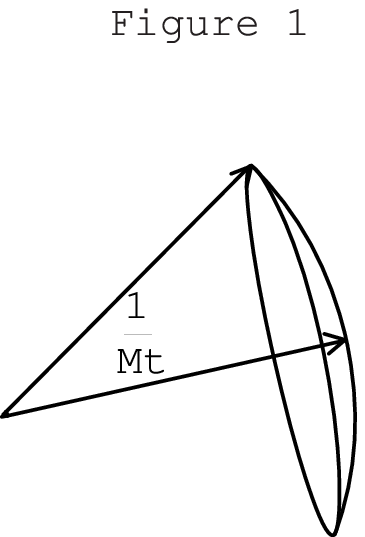}\end{figure}\begin{figure}[ptbptb]
\epsfbox{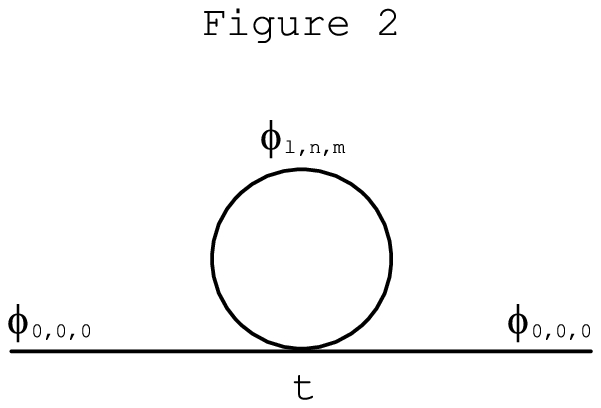}\end{figure}\begin{figure}[ptbptbptb]
\epsfbox{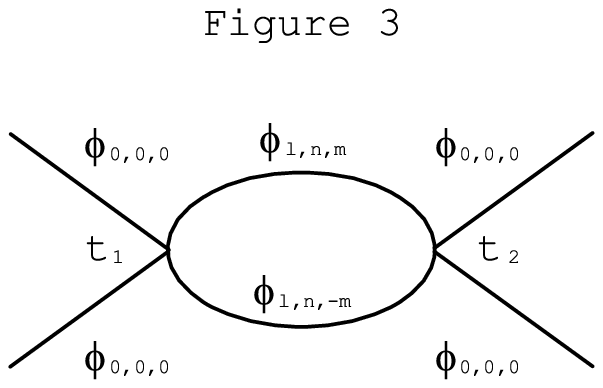}\end{figure}\begin{figure}[ptbptbptbptb]
\epsfbox{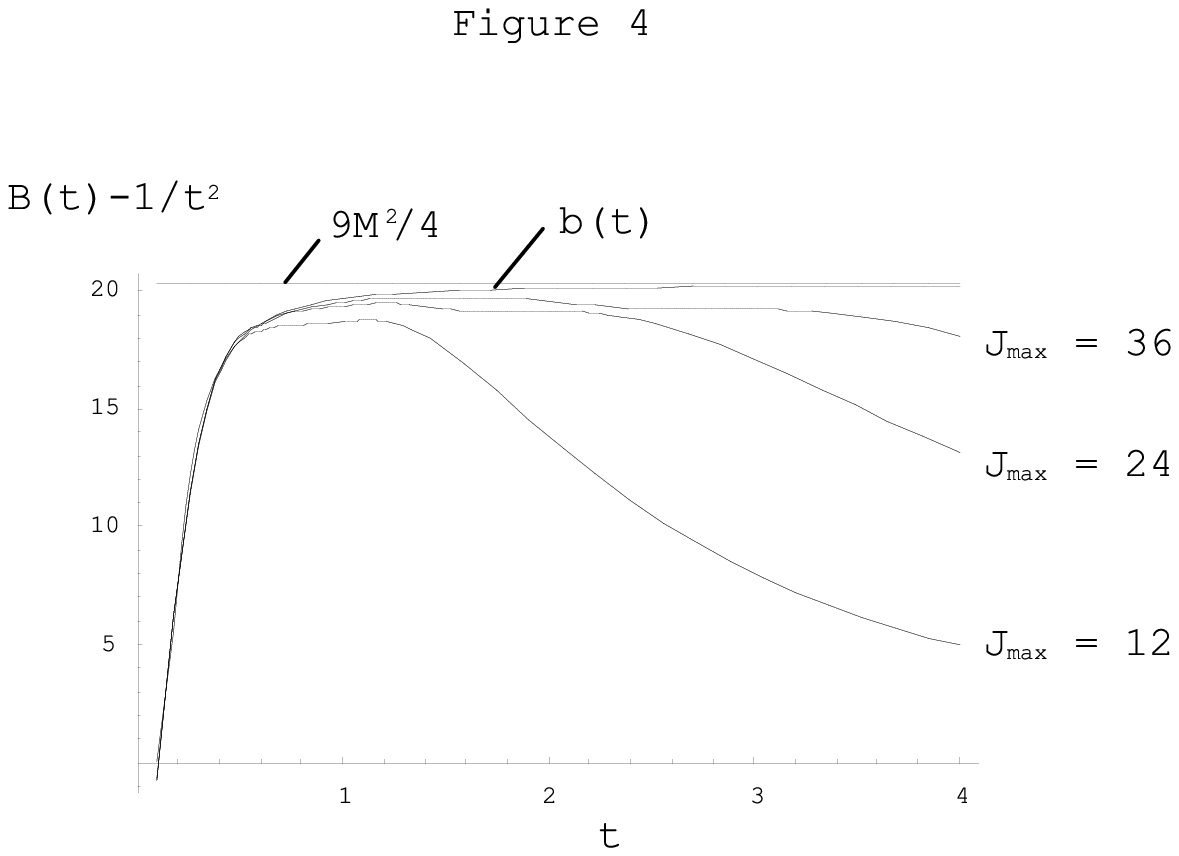}\end{figure}\begin{figure}[ptbptbptbptbptb]
\epsfbox{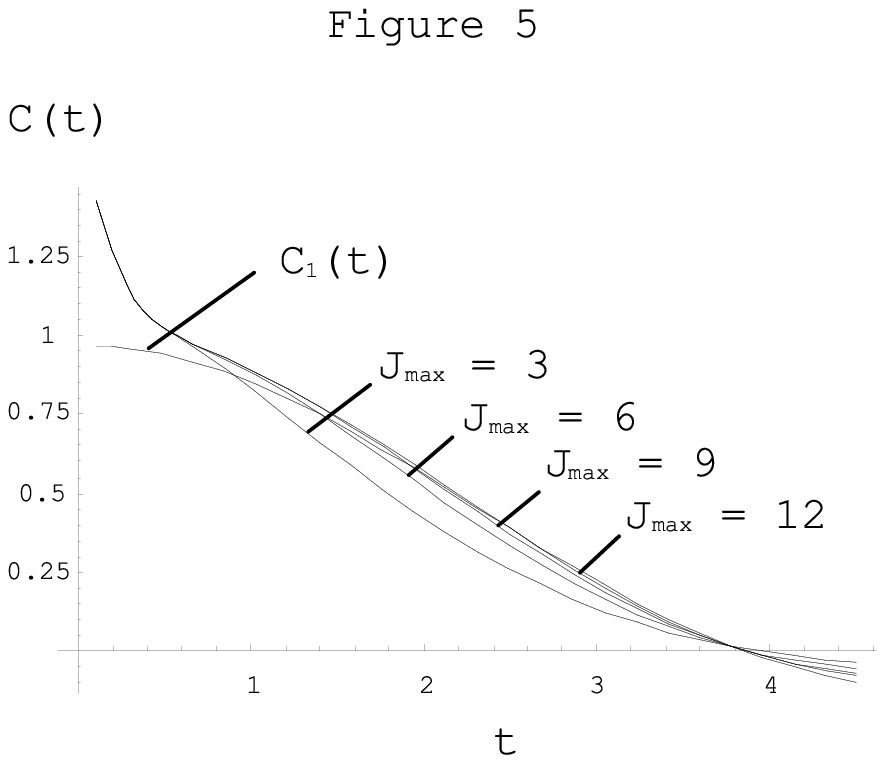}\end{figure}
\end{document}